\documentclass[preprint,aps,amsmath,nofootinbib,preprintnumbers,tightenlines]{revtex4}

\usepackage{epsfig}
\usepackage[OT2,T1]{fontenc}

\def\Bslash{B\!\!\!\!\slash}
\def\Dslash{D\!\!\!\!\slash}
\def\nslash{n\!\!\!\slash}
\def\bnslash{\bar n\!\!\!\slash}

\def\lslash{l\!\!\!\slash}

\newcommand{\ra}{\rightarrow}

\newcommand{\tr}{\mbox{tr}}

\newcommand{\cL}{{\cal L}}
\newcommand{\order}{{\cal O}}
\newcommand{\fr}{\frac}

\newcommand{\al}{\alpha}
\newcommand{\bt}{\beta}
\newcommand{\ga}{\gamma}
\newcommand{\de}{\delta}
\newcommand{\la}{\lambda}
\newcommand{\ep}{\epsilon}
\newcommand{\ve}{\varepsilon}

\newcommand{\Ga}{\Gamma}

\newcommand{\bn}{{\bar n}}

\def\bard{\textrm{\dj}}
\newcommand{\sdt}{\!\cdot\!}
\newcommand{\hbns}{\fr{\bnslash}{2}}
\newcommand{\hns}{\fr{\nslash}{2}}
\newcommand{\epsp}{\epsilon^\perp}
\newcommand{\hc}{\textrm{h.c.}}
\newcommand{\pqrn}{(p,\rho,A) \leftrightarrow (q,\nu,B)}
\newcommand{\cF}{{\cal F}}
\newcommand{\cJ}{{\cal J}}
\newcommand{\cP}{{\cal P}}
\newcommand{\lan}{\langle gg|}
\newcommand{\ran}{|0 \rangle}

\begin{document}

\preprint{ \vbox{
\hbox{arXiv:0808.3787}
\hbox{MIT-CTP-3976} }
}

\title{\boldmath
  The Chiral Anomaly in SCET
\vspace{0.4cm}
}

\author{Wouter J. Waalewijn}
\affiliation{Center for Theoretical Physics, Massachusetts Institute of
  Technology, Cambridge, MA 02139\vspace{0.3cm}}

\begin{abstract}
\vspace{0.3cm}
Anomalies are an infrared effect, but are often realized in effective theories in a non-trivial way. We study the chiral anomaly in Soft Collinear Effective Theory (SCET), where the anomaly equation has terms contributing at different orders in the power expansion. The chiral anomaly equations in SCET are computed up to NNLO in the power counting with external collinear and/or ultrasoft gluons. We do this by expanding the QCD anomaly equation, using the tree level (LO in $\al_s$) relations between QCD and SCET fields. The validity of this correspondence between the anomaly equations is confirmed by direct computation of the one-loop diagrams in SCET.
\end{abstract}

\maketitle

\newpage

\section{Introduction}

In calculations with a hierarchy of scales it is often useful to work with an effective field theory, in which ultraviolet degrees of freedom have been integrated out. Examples of this are Heavy Quark Effective Theory (HQET) and Soft Collinear Effective Theory (SCET). Or as a more well known example, integrating out a heavy quark.

Although anomalies \cite{Adler:1969gk, Bell:1969ts} come from ultraviolet divergences, they are actually an infrared effect because only massless particles contribute \cite{Frishman:1980dq}. If masses are generated through a Higgs mechanism as in the Standard Model, fields are massless above the symmetry breaking scale and can contribute to anomalies too.

Since they are an infrared effect one would expect effective field theories to reproduce the anomalies of the full theory, but this is not always a trivial matter. As an illustration we consider the case of integrating out a heavy quark doublet\footnote{In the Standard Model only the top would get integrated out before $W$ and $Z$ do. For simplicity we assume the existence of a heavy electroweak doublet.}. Simply removing it would spoil the anomaly cancelation for the gauge symmetries. D'Hoker and Farhi show in a detailed analysis that the effective action contains additional Wess-Zumino terms and terms involving the Goldstone-Wilczek current, that restore the gauge symmetry \cite{DHoker:1984ph, DHoker:1984pc}.

Another example is the axial anomaly in Chiral Perturbation Theory (ChPT). By itself ChPT would not include reactions such as $\pi^{0} \ra \ga \ga$ or $K \bar{K} \ra \pi^+ \pi^- \pi^0$. They do not show up because of the symmetry $M \ra -M$, where $M$ is the matrix of Goldstone bosons, which is not a symmetry of QCD. Again one needs to add Wess-Zumino terms and additional terms that couple to the gauge fields \cite{Wess:1971yu, Witten:1983tw}. \\

The goal of this paper is to study how the chiral anomaly is realized in SCET. SCET has been introduced to describe the dynamics of energetic light hadrons and provides a systematic way to separate the hard, collinear and soft scales \cite{Bauer:2000ew, Bauer:2000yr, Bauer:2001ct, Bauer:2001yt}. It captures the long distance physics in terms of collinear, soft and ultrasoft (usoft) fields. The short distance physics is absorbed into Wilson coefficients. So in SCET a quark field gets replaced by these different fields, which can each run around in loops. Currents in SCET often generate $\tfrac{1}{\ve^2}$ divergences at one-loop that are removed by renormalization. We will see that such divergences show up in individual SCET anomaly diagrams, but cancel when they are summed. Also in SCET there are vertices with two quark fields and more than one gluon, which lead to non-triangle anomaly diagrams. These graphs turn out to be important.

Which SCET fields one needs to consider, depends on the physical process one has in mind.
We will be looking at $\textrm{SCET}_\textrm{I}$ with one collinear direction $n^\mu$. The fields in this theory are collinear quarks $\xi_n$ and gluons $A_n^\mu$ and usoft quarks $q_{us}$ and gluons $A^\mu_{us}$. The momenta of the collinear fields scale like $p_c^\mu = (n \cdot p, \bn \cdot p, p_\perp) \sim Q(\lambda^2,1,\lambda)$ and for usoft fields as $p_{us}^\mu \sim Q \lambda^2$. Here $Q$ is the hard scale, $\lambda$ the SCET expansion parameter and $n^\mu$ and $\bn^\mu$ are light-cone basis vectors satisfying $n^2 = \bn^2 = 0$ and $n \sdt \bn = 2$. One can for example take $n^\mu = (1, 0, 0, 1)$ and $\bn^\mu = (1, 0, 0, -1)$. For a typical process $\lambda^2 = \Lambda_{QCD} / Q$.
We will also need the power counting of the fields, which are listed in table \ref{tab:pc_operators}.

To get a systematic expansion in the power counting, collinear fields have both a label momentum $p$ containing the collinear momentum and a (usoft) coordinate $x$, $\xi_{n,p}(x)$. The label momenta are picked out by the label operator $\cP$, e.g. $\cP_\perp^\mu \xi_{n,p} = p_\perp^\mu \xi_{n,p}$. The collinear covariant derivatives are then defined as
\begin{equation}
    i\bn \sdt D_c = \bn \sdt \cP + g\bn \sdt A_{n,q}, \qquad
    i D_c^{\perp\mu} = \cP_\perp^\mu + g A_{n,q}^{\perp \mu}, \qquad
    i n \sdt D = i n \sdt \partial + g n \sdt A_{n,q} + g n \sdt A_{us},
\end{equation}
and the usoft covariant derivative as
\begin{equation}
    iD_{us}^\mu = i\partial^\mu + g A_{us}^\mu.
\end{equation}
Note that these covariant derivatives are each homogenous in the power counting.
\begin{table}[t]
\begin{tabular}{|c|cccc|cc|cccc|}
    \hline
    \ & \multicolumn{4}{|c|}{collinear} & \multicolumn{2}{|c|}{usoft} & \multicolumn{4}{|c|}{covariant derivatives} \\ \hline
    operator \ & \ $\xi_n$ & $\bn \sdt A_n$ & $A_n^{\perp \mu}$ & $n \sdt A_n$ \ & \ $q_{us}$ & $A_{us}^\mu$ \ & \ $i\bn \sdt D_c$ & $iD_c^{\perp \mu}$ & $i n \sdt D$ & $iD_{us}^\mu$ \\
    \ power counting \ & $\lambda$ & $\lambda^0$ & $\lambda$ & $\lambda^2$ & $\lambda^3$ & $\lambda^2$ & $\lambda^0$ & $\lambda$ & $\lambda^2$ & $\lambda^2$ \\
    \hline
\end{tabular}
    \caption{Power counting of the various fields and operators in SCET}
    \label{tab:pc_operators}
\end{table}
We are now ready to write down the leading order Lagrangian for a collinear quark $\xi_n$ \cite{Bauer:2000yr}
\begin{equation}
    \cL^{(0)}_{\xi\xi} = \bar{\xi}_n(x) \left( in \sdt D + i \Dslash^c_\perp \frac{1}{i\bn \sdt D^c} \ i\Dslash^c_\perp \right) \hbns \ \xi_n(x),
\end{equation}
where it is understood that the (suppressed) label momenta are summed over and that the label momenta of each term in $\cL$ is conserved. \\

This paper is organized as follows: we start in section \ref{sec:match_QCD_anomaly} by matching the chiral anomaly equation in QCD onto SCET, using the tree level relations between fields. For simplicity we first restrict ourselves to LO in the power counting, postponing the derivation up to NNLO to section \ref{sec:match_QCD_anomaly_higher}. Whenever we speak of e.g. LO and NLO in this paper, we will always be referring to the order in the power expansion $\lambda$ and not to radiative corrections. No contributions are expected beyond one-loop due to the Adler and Bardeen theorem \cite{Adler:1969er}.

We then verify these SCET anomaly equations by computing the chiral anomaly from one-loop graphs in SCET. This is done at LO in sections \ref{sec:LO_nbar}-\ref{sec:LO_general}, NLO in section \ref{sec:NLO_n} and NNLO in section \ref{sec:SCET_anomaly_NNLO}. At LO only collinear fields play a role and so we expect agreement, since SCET with only collinear fields is just a boosted version of QCD.
Things become more interesting at higher orders when diagrams with both collinear and usoft fields show up, but we find that the derived SCET anomaly equations are correct.
We end with some concluding remarks in section \ref{sec:conclusions} and list the full SCET anomaly equations up to NLO. The appendix contains a table of the loop integrals we used.

\section{The SCET anomaly at LO}
\label{sec:SCET_anomaly_LO}

\subsection{Matching the QCD anomaly onto SCET}
\label{sec:match_QCD_anomaly}

In this section we will calculate the chiral anomaly in SCET at LO. We start by expanding the left- and righthand side of the QCD anomaly equation
\begin{equation}\label{anomaly_in_QCD}
    \partial_\mu J_5^\mu = -\fr{g^2}{16\pi^2} \ \epsilon_{\alpha\beta\mu\nu} \tr[F^{\alpha\beta}F^{\mu\nu}]
\end{equation}
in the power counting, where $J_5^\mu = \bar{\Psi} \ga^\mu \ga_5 \Psi$.
We use the tree level relations between fields, effectively matching QCD onto SCET at tree level. In the next subsections we will explicitly check that the SCET anomaly equation we find, holds at one-loop. \\

We introduce the following notation for the left- and righthand side of equation \eqref{anomaly_in_QCD}
\begin{equation}
    \cJ \equiv \partial_\mu J_5^\mu, \qquad \cF \equiv -\fr{g^2}{16\pi^2} \ \epsilon_{\alpha\beta\mu\nu} \tr[F^{\alpha\beta}F^{\mu\nu}],
\end{equation}
and write their expansions as
\begin{eqnarray}\label{J_F_expansion}
    \cJ &=& \cJ^{(4)} + \cJ^{(5)} + \cJ^{(6)} + \dots, \\
    \cF &=& \cF^{(4)} + \cF^{(5)} + \cF^{(6)} + \dots, \nonumber
\end{eqnarray}
where the order in $\la$ is indicated in brackets.
When calculating the one loop anomaly diagrams, we will (as usual) restrict ourselves to two outgoing gluons. Gauge invariance then determines anomaly matrix elements for more gluons.
Let $p$, $q$ be the momenta of these two gluons. For simplicity we assume that they have no component in the $\perp$ direction, $p_\perp = q_\perp = 0$. We will also assume that the gluons are $\perp$-polarized.
The matrix element of $\cJ$ is then given by
\begin{equation}\label{deriv_J5}
    \lan \cJ \ran = \tfrac{1}{2}i n \sdt (p+q) \ \lan \bn \sdt J_5 \ran + \tfrac{1}{2}i \bn \sdt (p+q) \ \lan n \sdt J_5 \ran.
\end{equation}
The tree level relation between the QCD and SCET (collinear) field is \cite{Bauer:2000yr}
\begin{equation}\label{J5mu_psi_in_SCET}
    \Psi = \underbrace{\vphantom{\fr{1}{i \bn \sdt D_c}} \xi_n}_\la + \underbrace{\fr{1}{i \bn \sdt D_c} \ i\Dslash_\perp^c \hbns \ \xi_n}_{\la^2} + \dots,
\end{equation}
with the power counting as indicated. This leads to
\begin{equation}\label{barn_J5_LO}
    \bn \sdt J_5^{(2)} = \bar{\xi}_n \bnslash \ga_5 \xi_n.
\end{equation}
For $n \sdt J_5$ we get an $\nslash$. But $\nslash \xi_n = 0$, so the lowest non-vanishing contribution is at $\order(\lambda^4)$
\begin{eqnarray}\label{n_J5_LO} \nonumber
    n \sdt J_5^{(4)} &=&
    \left(\fr{1}{i\bn \sdt D_c} \ i\Dslash_\perp^c \hbns \ \xi_n \right)^\dagger
    \ga^0 \nslash \ga_5
    \left(\fr{1}{i\bn \sdt D_c} \ i\Dslash_\perp^c \hbns \ \xi_n \right) \\
    &=& -\bar{\xi}_n i \! \overleftarrow{\Dslash}_\perp^c \fr{1}{i\bn \sdt \overleftarrow{D}_c}
    \ \bnslash \ga_5 \fr{1}{i\bn \sdt D_c} \ i\Dslash_\perp^c \xi_n.
\end{eqnarray}
However, it is important to note that $n \sdt J_5^{(4)}$ contributes at the same order as the $\bn \sdt J_5$ term in the anomaly equation because of the power counting for collinear momenta,
\begin{equation}\label{J_at_O_4}
    \lan\cJ^{(4)}\ran =  \tfrac{1}{2}i n \sdt (p+q) \ \lan \bn \sdt J_5^{(2)} \ran + \tfrac{1}{2}i \bn \sdt (p+q) \ \lan n \sdt J_5^{(4)} \ran.
\end{equation}

We will first study these terms separately by imposing $\bn \sdt (p+q)=0$ or $n \sdt (p+q)=0$ in the following subsections. With these additional assumptions we have to keep gluons off- shell, $p^2, q^2 \neq 0$, because everything would otherwise trivially vanish. At the end we will drop these additional assumptions, effectively combining the two cases. \\

Now we consider the righthand side of the anomaly equation. With our assumptions the important terms are
\begin{eqnarray}\label{D_mu}
    i n \sdt D &=& i n \sdt \partial + g n \sdt A_{n,q} + g n \sdt A_{us} = \underbrace{i n \sdt \partial}_{\la^2} + \dots, \\
    i \bn \sdt D &=& i \bn \sdt D_c + W i \bn \sdt D_{us} W^\dagger = \underbrace{\bn \sdt \cP}_1 + \underbrace{i \bn \sdt \partial}_{\la^2} + \dots, \nonumber \\
    i D_\perp^\mu &=& i D_c^{\perp \mu} + W iD_{us}^{\perp \mu} W^\dagger = \underbrace{g A_{n,q}^{\perp \mu}}_\la + \underbrace{\vphantom{g A_{n,q}^{\perp \mu}} g A_{us}^{\perp \mu}}_{\la^2} + \dots \nonumber
\end{eqnarray}
$W$ is the collinear Wilson line that only contains the $\bn \sdt A_n$ component of the gluon field
\begin{equation}
    W = 1 - g \ \frac{1}{\bn \sdt \cP} \ \bn \sdt A_n + \dots
\end{equation}
and can be dropped because we are only looking at $\perp$-polarized gluons.
Writing
\begin{equation}\label{FFdual}
    \epsilon_{\al\bt\mu\nu} F^{\al\bt} F^{\mu\nu} =
    -\fr{4}{g^2} \ \epsilon_{\al\bt\mu\nu} iD^\al iD^\bt iD^\mu iD^\nu,
\end{equation}
the only contribution we need to consider at $\order(\la^4)$ is
\begin{eqnarray}
    \cF^{(4)} &=& -\fr{g^2}{16\pi^2} \times -\fr{4}{g^2} \ \epsilon_{\al\mu\bt\nu} \fr{\bn^\al}{2} \fr{n^\bt}{2}
    \ \tr \big[i n \sdt \partial \ g A_{n,p}^{\perp\mu} \ \bn \sdt \cP \ g A_{n,q}^{\perp\nu}
    -i n \sdt \partial \ g A_{n,q}^{\perp\nu} \ \bn \sdt \cP \ g A_{n,p}^{\perp\mu} \nonumber \\
    & \ & \hspace{4cm} - \bn \sdt \cP \ g A_{n,p}^{\perp\mu} \ i n \sdt \partial \ g A_{n,q}^{\perp\nu}
    + \bn \sdt \cP \ g A_{n,q}^{\perp\nu} \ i n \sdt \partial \ g A_{n,p}^{\perp\mu}\big].
\end{eqnarray}
The corresponding matrix element is given by
\begin{eqnarray}\label{FFdual_at_lambda_4}
    \lan \cF^{(4)} \ran =
    -\fr{g^2}{8\pi^2} \ (n \sdt p \ \bn \sdt q - n \sdt q \ \bn \sdt p) \de^{AB} \epsp_{\mu\nu} \ep^{\mu *}_A(p) \ep^{\nu *}_B(q),
\end{eqnarray}
where $\ep^{\mu *}_A(p)$, $\ep^{\nu *}_B(q)$ are the polarization vectors of the gluons and we assume the generators are normalized as $\tr[T^A T^B]=\tfrac{1}{2} \de^{AB}$.
The two-dimensional $\epsp$ is defined as
\begin{equation}
    \epsp_{\mu \nu} \equiv
    \tfrac{1}{2} \epsilon_{\al \bt \mu \nu} \bn^\al n^\bt =
    \tfrac{1}{8} i \ \tr[\ga_5 \bnslash \nslash \ga^\perp_\mu \ga^\perp_\nu].
\end{equation}

Having derived the explicit form of the SCET anomaly equation at LO
\begin{equation}
    \cJ^{(4)} = \cF^{(4)},
\end{equation}
we will verify this result by calculating the one loop diagrams.

\subsection{LO with $\bn \sdt (p+q)=0$}
\label{sec:LO_nbar}

\begin{figure*}[t]
  \centerline{
   \mbox{\epsfxsize=4truecm \hbox{\epsfbox{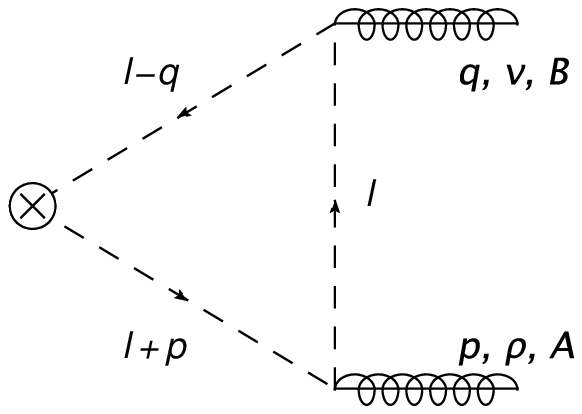}} } \qquad
   \mbox{\epsfxsize=4truecm \hbox{\epsfbox{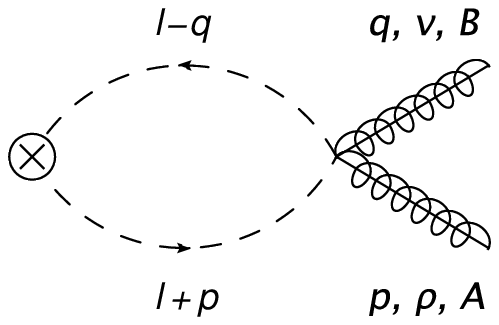}} }}
  \caption{Triangle and bubble diagram contributing to the anomaly at LO, for $\bn \sdt (p+q)=0$}
   \label{fig:LOnbar}
\end{figure*}

We start by considering the simplest case, namely $\bn \sdt (p+q)=0$ for the momenta of the external gluons. The axial current is then given by
\begin{equation}
    \lan \cJ^{(4)} \ran = \tfrac{1}{2}i n \sdt (p+q) \ \lan \bn \sdt J_5^{(2)} \ran
    = \tfrac{1}{2}i n \sdt (p+q) \ \lan \bar{\xi}_n \bnslash \ga_5 \xi_n \ran
\end{equation}
and there are two diagrams that contribute at this order, shown in figure \ref{fig:LOnbar}.
As mentioned before, taking the gluons on-shell would make everything vanish because of $\bn \sdt (p+q)=0$. We therefore keep the gluons off-shell, which generally takes care of any IR divergences as well. Because we assume $\bn \sdt (p+q)=0$, we still need to be careful when combining denominators as $(l+p)^2$ and $(l-q)^2$. We will elaborate on this for the bubble diagram. \\

In dimensional regularization with $d=4-2\ve$, the triangle diagram is given by
\begin{eqnarray}\label{triagle_LO_nbar}
    A_T = \tfrac{1}{2} i n \sdt (p+q) \ (-1) \ \int \bard l \ \tr \Bigg[\bnslash \gamma_5 \ i \hns \fr{\bn \sdt (l-q)}{(l-q)^2} \ ig T^B \left(\fr{\ga^\perp_\nu \lslash_\perp}{\bn \sdt l} + \fr{\lslash_\perp \ga^\perp_\nu}{\bn \sdt (l-q)}\right) \hbns \
    i\hns \fr{\bn \sdt l}{l^2} \nonumber \\
    \times \ ig T^A \left(\fr{\ga^\perp_\rho \lslash_\perp}{\bn \sdt (l+p)} + \fr{\lslash_\perp \ga^\perp_\rho}{\bn \sdt l}\right) \hbns \
    i\hns \fr{\bn \sdt (l+p)}{(l+p)^2}\Bigg] + \pqrn, \
\end{eqnarray}
where $\int \bard l = \int d^d l (2\pi)^{-d}$. We will always suppress polarization vectors.
To avoid any complications related to $\ga_5$ in dim. reg., we never (anti)commute with $\ga_5$ in our calculations.
Since the external momenta have no $\perp$-component, we can immediately replace $l_\perp^\al l_\perp^\bt \ra \tfrac{1}{\Delta} l_\perp^2 g_\perp^{\al\bt}$, where $\Delta=d-2=2-2\ve$. We then use
\begin{eqnarray}\label{gamma_identities}
    \ga_\perp^\mu \ga^\perp_\mu = \Delta, \qquad
    \ga_\perp^\mu \ga^\perp_\nu \ga^\perp_\mu = -(\Delta-2) \gamma^\perp_\nu = 2 \ve \gamma^\perp_\nu, \\
    \ga_\perp^\mu \ga^\perp_\nu \ga^\perp_\rho \ga^\perp_\mu = 4 g^\perp_{\nu \rho} - (4 - \Delta) \ga^\perp_\nu \ga^\perp_\rho. \qquad \nonumber
\end{eqnarray}
This leads to
\begin{eqnarray}
    A_T &=& -ig^2 n \sdt (p+q) \delta^{AB} \epsp_{\rho \nu} (I_1 + \ve I_2 + \ve I_3 - (1+2\ve) I_4) + \pqrn \nonumber \\
    &=& \frac{g^2}{8\pi^2} \ n \sdt (p+q) \bn \sdt p \ \delta^{AB} \epsp_{\rho \nu},
\end{eqnarray}
where
\begin{eqnarray}\label{integrals_triangle_LO_nbar}
    I_1 = \int \bard l \fr{\bn \sdt (l-q) \ \bn \sdt (l+p) \ l_\perp^2}{\bn \sdt l \ (l-q)^2 \ l^2 \ (l+p)^2}, \qquad
    I_2 = \int \bard l \fr{\bn \sdt (l-q) \ l_\perp^2}{(l-q)^2 \ l^2 \ (l+p)^2}, \\
    I_3 = \int \bard l \fr{\bn \sdt (l+p) \ l_\perp^2}{(l-q)^2 \ l^2 \ (l+p)^2}, \qquad
    I_4 = \int \bard l \fr{\bn \sdt l \ l_\perp^2}{(l-q)^2 \ l^2 \ (l+p)^2}. \nonumber
\end{eqnarray}
We computed these integrals using an appropriate Feynman parametrization. A table with all the necessary integrals can be found in the appendix. \\

The contribution from the bubble diagram (right panel of fig. \ref{fig:LOnbar}) is
\begin{eqnarray}\label{bubble_LO_nbar}
    A_B &=& \tfrac{1}{2} i n \sdt (p+q) \ (-1) \int \bard l \ \tr \Bigg[ \bnslash \gamma_5 \ i \hns \fr{\bn \sdt (l-q)}{(l-q)^2} \nonumber \\
    & \ & \hspace{4cm}
    \times \ ig^2 \left( \fr{T^A T^B \ga^\perp_\rho \ga^\perp_\nu}{\bn \sdt (l+p-q)} + \fr{T^B T^A \ga^\perp_\nu \ga^\perp_\rho}{\bn \sdt l} \right) \hbns \ i \hns \fr{\bn \sdt (l+p)}{(l+p)^2} \Bigg] \nonumber \\
    &=& ig^2 n \sdt (p+q) \delta^{AB} \epsp_{\rho \nu} \int \bard l \left(\fr{1}{\bn \sdt l} - \fr{1}{\bn \sdt (l+p-q)}\right) \fr{\bn \sdt (l-q) \ \bn \sdt (l+p)}{(l-q)^2 \ (l+p)^2}.
\end{eqnarray}
By sending $l \ra -l-p+q$ we see that the two terms are the same (giving a factor of 2, rather than canceling each other). We do not have a second contribution from $\pqrn$ for this diagram.
As mentioned before, we need to be careful in dealing with IR divergences here. Writing $\delta = \bn \sdt (p+q) \ra 0$, we encounter $\delta^{-\ve}$ when doing this integral. Obviously the order of taking $\delta \ra 0$ and $\ve \ra 0$ matters and we should first take $\ve \ra 0$ with $\delta \neq 0$ to regulate any IR divergences. Proceeding along this way, we still find $A_B = 0$. \\

We now add $A_T$ and $A_B$ and compare with $\lan \cF^{(4)} \ran$ as given in \eqref{FFdual_at_lambda_4}. They agree. This was expected since at this order only collinear fields are involved, which is just a boosted version of QCD.

\subsection{LO with $n \sdt (p+q)=0$}
\label{sec:LO_n}

\begin{figure*}[b]
  \centerline{
   \mbox{\epsfxsize=2truecm \hbox{\epsfbox{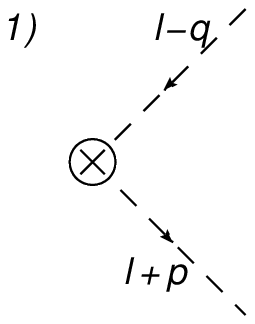}} } \qquad
   \mbox{\epsfxsize=2.5truecm \hbox{\epsfbox{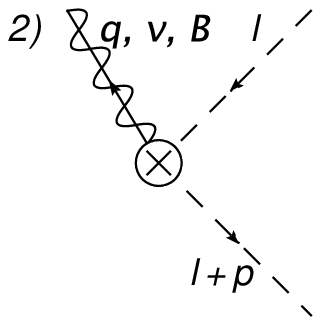}} } \qquad
   \mbox{\epsfxsize=3truecm \hbox{\epsfbox{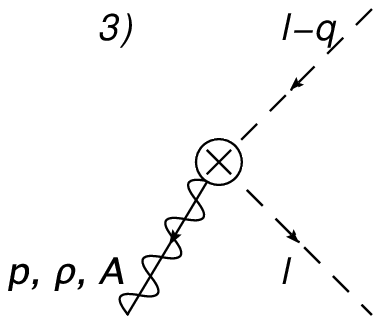}} } \qquad
   \mbox{\epsfxsize=3truecm \hbox{\epsfbox{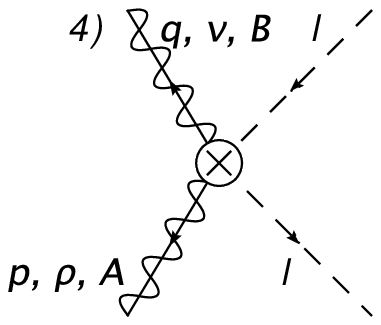}} }}
  \caption{$n \sdt J_5^{(4)}$ can be inserted into diagrams in various ways}
   \label{fig:nJ4}
\end{figure*}

We move on to the next case: $n \sdt (p+q)=0$.
This time the relevant term of the axial current comes from $n \sdt J_5^{(4)}$
\begin{equation}
    \lan\cJ^{(4)}\ran = \tfrac{1}{2}i \bn \sdt (p+q) \ \lan
    -\bar{\xi}_n i \! \overleftarrow{\Dslash}_\perp^c \fr{1}{i\bn \sdt \overleftarrow{D}_c}
    \bnslash \ga_5 \fr{1}{i\bn \sdt D_c} \ i\Dslash_\perp^c \xi_n \ran.
\end{equation}
It can be part of a diagram in various different ways, for example a gluon can now come out of the current. These different ways are pictured in fig. \ref{fig:nJ4} and the corresponding expressions are
\begin{eqnarray}\label{nJ_LO_n_rules}
    \textrm{1)} \ -\bar{\xi}_n \frac{\lslash_\perp}{\bn \sdt (l+p)} \bnslash \gamma_5 \frac{\lslash_\perp}{\bn \sdt (l-q)} \ \xi_n, \qquad
    \textrm{2)} \ -\bar{\xi}_n \frac{\lslash_\perp}{\bn \sdt (l+p)} \bnslash \gamma_5 \frac{g T^B \gamma^\perp_\nu}{\bn \sdt (l-q)} \ \xi_n, \\
    \textrm{3)} \ -\bar{\xi}_n \frac{g T^A \gamma^\perp_\rho}{\bn \sdt (l+p)} \bnslash \gamma_5 \frac{\lslash_\perp}{\bn \sdt (l-q)} \ \xi_n, \qquad
    \textrm{4)} \ -\bar{\xi}_n \frac{g T^A \gamma^\perp_\rho}{\bn \sdt (l+p)} \bnslash \gamma_5 \frac{g T^B \gamma^\perp_\nu}{\bn \sdt (l-q)} \ \xi_n. \nonumber
\end{eqnarray}
\begin{figure*}[t]
  \centerline{
   \mbox{\epsfxsize=4truecm \hbox{\epsfbox{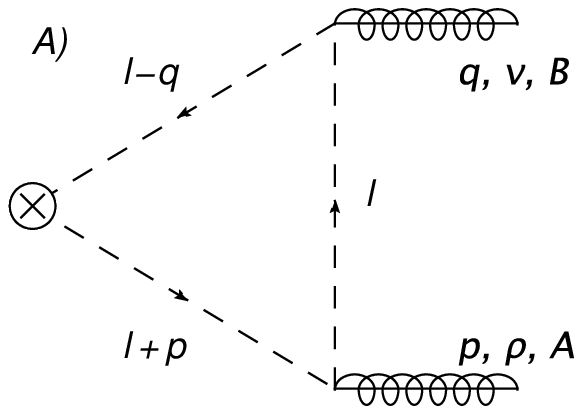}} } \qquad
   \mbox{\epsfxsize=4truecm \hbox{\epsfbox{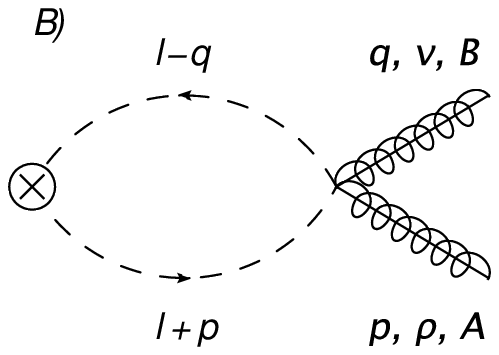}} }}
  \vspace{0.5cm}
  \centerline{
   \mbox{\epsfxsize=5truecm \hbox{\epsfbox{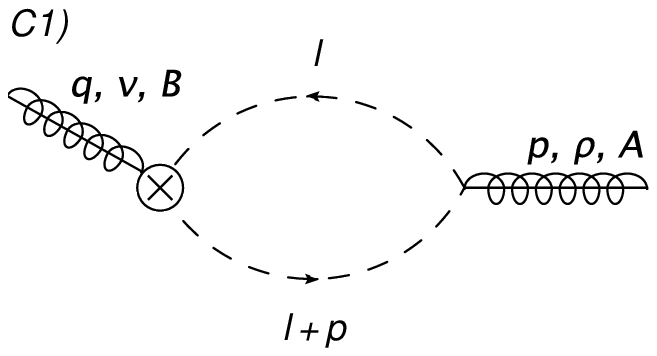}} } \qquad
   \mbox{\epsfxsize=5truecm \hbox{\epsfbox{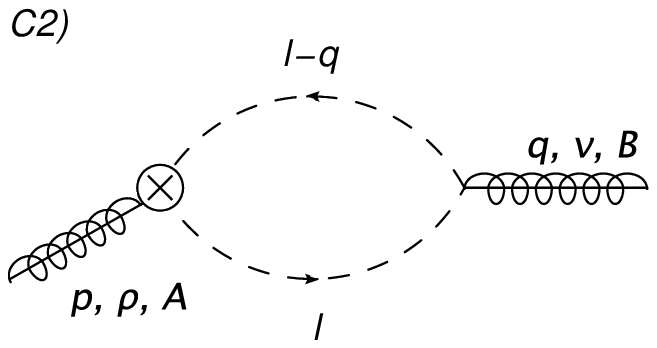}} } \qquad
   \mbox{\epsfxsize=3.8truecm \hbox{\epsfbox{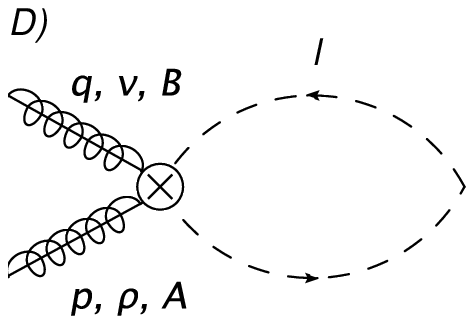}} }}
  \caption{Diagrams contributing to the anomaly at LO for $n \sdt (p+q)=0$}
   \label{fig:LOn}
\end{figure*}

First of all we get a triangle fig. \ref{fig:LOn}A and bubble diagram fig. \ref{fig:LOn}B coming from 1)
\begin{eqnarray}\label{LO_n_rule_a}
    \tilde{A}_A + \tilde{A}_B &=& ig^2 \bn \sdt (p+q) \delta^{AB} \epsp_{\rho \nu}
    ((1+2\ve) \tilde{I}_1 - \ve\tilde{I}_2 - \ve\tilde{I}_3 - \tilde{I}_4 - (1+2\ve)\tilde{I}_5) \nonumber \\
    & \ & + \pqrn,
\end{eqnarray}
where
\begin{eqnarray}\label{integrals_LO_n}
    \tilde{I}_1 = \int \bard l \fr{l_\perp^4}{\bn \sdt l \ (l-q)^2 \ l^2 \ (l+p)^2}, & \ &
    \tilde{I}_2 = \int \bard l \fr{l_\perp^4}{\bn \sdt (l+p) \ (l-q)^2 \ l^2 \ (l+p)^2}, \\
    \tilde{I}_3 = \int \bard l \fr{l_\perp^4}{\bn \sdt (l-q) \ (l-q)^2 \ l^2 \ (l+p)^2}, & \ &
    \tilde{I}_4 = \int \bard l \fr{\bn \sdt l \ l_\perp^4}{\bn \sdt (l-q) \ \bn \sdt (l+q) \ (l-q)^2 \ l^2 \ (l+p)^2}, \nonumber \\
    \tilde{I}_5 = \int \bard l \fr{l_\perp^2}{\bn \sdt l \ (l-q)^2 \ (l+p)^2}. & \ & \nonumber
\end{eqnarray}
We also have diagrams fig. \ref{fig:LOn}C1 with 2) and fig. \ref{fig:LOn}C2 with 3), where one gluon comes out of the current
\begin{equation}\label{LO_n_rule_b_c}
    \tilde{A}_{C1} + \tilde{A}_{C2} = 2 ig^2 \bn \sdt(p+q) \delta^{AB} \epsp_{\rho \nu} (\tilde{I}_6 + \ve \tilde{I}_7) + \pqrn,
\end{equation}
with
\begin{equation}
    \tilde{I}_6 = \int \bard l \fr{\bn \sdt l \ l_\perp^2}{\bn \sdt (l-q) \ \bn \sdt (l+p) \ l^2 \ (l+p)^2}, \qquad
    \tilde{I}_7 = \int \bard l \fr{l_\perp^2}{\bn \sdt (l-q) \ l^2 \ (l+p)^2}.
\end{equation}
Finally there is also a diagram fig. \ref{fig:LOn}D where both gluons come out of the current, which involves 4). This turns out to be zero, $\tilde{A}_D = 0$. \\

We simplify the remaining integrals using
\begin{equation}\label{simplify_integrals}
    l_\perp^2 = l^2 - \bn \sdt l \ n \sdt l = (l-q)^2 - \bn \sdt (l-q) \ n \sdt (l-q) = (l+p)^2 - \bn \sdt (l+p) \ n \sdt (l+p),
\end{equation}
where $l_\perp^2 = -\vec{l}_\perp^2$ if $g^{\mu \nu} = \textrm{diag}(+---)$.
For example
\begin{equation}\label{integral_2_tilde}
    \tilde{I}_2 = \int \bard l \left(\fr{l_\perp^2}{\bn \sdt (l+p) \ (l-q)^2 \ l^2} -
    \fr{n \sdt (l+p) \ l_\perp^2}{(l-q)^2 \ l^2 \ (l+p)^2} \right) = \tilde{I}_{2A} + \tilde{I}_{2B}.
\end{equation}
Doing the math, we find
\begin{equation}\label{anomaly_LO_n_result}
    \lan \cJ^{(4)} \ran = \tilde{A}_A + \tilde{A}_B + \tilde{A}_{C1} + \tilde{A}_{C2} + \tilde{A}_D = -\frac{g^2}{8\pi^2} \ \bn \sdt (p+q) n \sdt p \ \delta^{AB} \epsp_{\rho \nu}.
\end{equation}
Again this agrees with $\lan \cF^{(4)} \ran$, as expected because the calculation so far only involved collinear fields.

\subsection{LO general case}
\label{sec:LO_general}

Finally we now drop the additional assumptions (e.g. $n \sdt (p+q)=0$) and study the general case. These assumptions allowed us to look at one term of $\lan \cJ^{(4)} \ran$ in \eqref{J_at_O_4} at the time, so we basically have to add the anomalies of sections \ref{sec:LO_nbar} and \ref{sec:LO_n}. The only other place these assumptions were used, is in simplifying the Feynman integrals. We can now take the gluons on shell and we will assume
\begin{equation}
    n \sdt p = 0, \qquad \bn \sdt q = 0.
\end{equation}
After repeating the above analysis and performing the Feynman integrals we find
\begin{equation}\label{anomaly_LO_general_result}
    \lan \cJ^{(4)} \ran = \fr{g^2}{8\pi^2} \ \bn \sdt p \ n \sdt q \ \delta^{AB} \epsp_{\rho \nu} = \lan \cF^{(4)} \ran.
\end{equation}
Thus the SCET anomaly equation at LO $\cJ^{(4)}=\cF^{(4)}$, which we derived by expanding, is correct at one-loop.
In the previous sections the bubble diagrams in fig. \ref{fig:LOn}C1, C2 contributed, but the bubble diagram with the two gluon vertex in fig. \ref{fig:LOnbar} and fig. \ref{fig:LOn}B turned out to be zero. However, in this general case with on-shell momenta we do get a genuine contribution from them.

\section{The SCET anomaly at NLO}
\label{sec:SCET_anomaly_NLO}

\subsection{Matching the QCD anomaly onto SCET at higher orders}
\label{sec:match_QCD_anomaly_higher}

We will now move on to calculating the chiral anomaly in SCET beyond LO in the power counting, by taking into account the higher order terms in eq. \eqref{J_F_expansion}.
Again we start by match- ing the QCD anomaly equation onto SCET at tree level. The derivation is similar to that in section \ref{sec:match_QCD_anomaly}, but this time we need the tree level relation between the QCD and SCET fields to higher order in $\lambda$. Using the result from \cite{Bauer:2003mga} we have
\begin{equation}\label{psi_in_SCET_full}
    \Psi = \underbrace{\vphantom{\fr{1}{i \bn \sdt D_c}} \xi_n}_\la + \underbrace{\vphantom{\fr{1}{i \bn \sdt D_c}} W q_{us}}_{\la^3} + \underbrace{\fr{1}{i \bn \sdt D_c} \ i\Dslash_\perp^c \hbns \ \xi_n}_{\la^2} + \underbrace{\vphantom{\fr{1}{i \bn \sdt D_c}} W \fr{1}{\bn \sdt \cP} \ i\Dslash_\perp^{us} \hbns \ W^\dagger \xi_n}_{\la^3} + \dots
\end{equation}
We find
\begin{eqnarray}\label{barn_J5}
    \bn \sdt J_5^{(2)} &=& \bar{\xi}_n \bnslash \ga_5 \xi_n, \\
    \bn \sdt J_5^{(3)} &=& 0, \\
    \bn \sdt J_5^{(4)} &=& \bar{\xi}_n \bnslash \ga_5 W q_{us} + \hc,
\end{eqnarray}
and
\begin{eqnarray}\label{n_J5}
    n \sdt J_5^{(4)} &=&
    -\bar{\xi}_n i \! \overleftarrow{\Dslash}_\perp^c \fr{1}{i\bn \sdt \overleftarrow{D}_c} \ \bnslash \ga_5 \fr{1}{i\bn \sdt D_c} \ i\Dslash_\perp^c \xi_n, \\
    n \sdt J_5^{(5)} &=&
    -2 \bar{\xi}_n i \! \overleftarrow{\Dslash}_\perp^c \fr{1}{i\bn \sdt \overleftarrow{D}_c} \ \ga_5 (W q_{us} + W \fr{1}{\bn \sdt \cP} \ i\Dslash_\perp^{us} \hbns \ W^\dagger \xi_n) + \hc
\end{eqnarray}
Expanding the other side of the QCD anomaly equation gives
\begin{eqnarray}
    \label{FFdual_at_lambda_4_2}
    \lan \cF^{(4)} \ran &=& -\fr{g^2}{8\pi^2} \ (n \sdt p \ \bn \sdt q - n \sdt q \ \bn \sdt p) \de^{AB} \epsp_{\mu\nu} \ep^{\mu *}_A(p) \ep^{\nu *}_B(q), \\
    \label{FFdual_at_lambda_5}
    \lan \cF^{(5)} \ran &=& \fr{g^2}{8\pi^2} \ \bn \sdt p \ n \sdt q \ \de^{AB} \epsp_{\mu\nu} \ep^{\mu *}_A(p) \ep^{\nu *}_B(q), \\ \label{FFdual_at_lambda_6}
    \lan \cF^{(6)} \ran &=& -\fr{g^2}{8\pi^2} \ (n \sdt p \ \bn \sdt q_r - n \sdt q \ \bn \sdt p_r) \de^{AB} \epsp_{\mu\nu} \ep^{\mu *}_A(p) \ep^{\nu *}_B(q),
\end{eqnarray}
where the subscript in $p_r$ denotes residual momentum, $p_r^\mu \sim \la^2$.
We already checked at LO that the SCET anomaly equation found this way is correct, $\cJ^{(4)}=\cF^{(4)}$. In the remainder of this paper we will do the NLO and NNLO calculations. \\

When we go beyond LO, we obviously get contributions to the anomaly from higher order currents such as $n \sdt J_5^{(5)}$. However there are also contributions from diagrams involving the leading current and insertion(s) of subleading Lagrangians. To be specific, the SCET anomaly equation at NLO and NNLO read
\begin{align}
    \label{NLOanomaly}
    &\textrm{T} \big\{ \cJ^{(4)} \ i\cL^{(1)} \big\} + \cJ^{(5)} =
    \textrm{T} \big\{ \cF^{(4)} \ i\cL^{(1)} \big\} +\cF^{(5)}, \\
    \label{NNLOanomaly}
    &\textrm{T} \big\{ \cJ^{(4)} \ \tfrac{1}{2} \ i\cL^{(1)} \ i\cL^{(1)} \big\} + \textrm{T} \big\{ \cJ^{(4)} \ i\cL^{(2)} \big\} + \textrm{T} \big\{ \cJ^{(5)} \ i\cL^{(1)} \big\} + \cJ^{(6)} = \nonumber \\
    &\hspace{3cm} \textrm{T} \big\{ \cF^{(4)} \ \tfrac{1}{2} \ i\cL^{(1)} \ i\cL^{(1)} \big\} + \textrm{T} \big\{ \cF^{(4)} \ i\cL^{(2)} \big\} + \textrm{T} \big\{ \cF^{(5)} \ i\cL^{(1)} \big\} + \cF^{(6)},
\end{align}
where $\textrm{T} \big\{ \cJ^{(4)} \ i\cL^{(1)} \big\} = \int d^4x \ \textrm{T} \ \cJ^{(4)}(0) \ i\cL^{(1)}(x)$ etc. $\cL^{(1)}$ and $\cL^{(2)}$ refer to all terms in the quark and gluon action of the respective order.

Let us consider the righthand side of these equations. Calculating the anomaly at one-loop corresponds to tree level diagrams for $\cF^{(i)}$. The only non-trivial diagrams come from insertions on external gluon lines. If one would include insertions on external gluon lines you would get contributions on both sides of these equations, which are equal because these insertions are not part of the loop. We will not consider such diagrams and so the right of \eqref{NLOanomaly} and \eqref{NNLOanomaly} reduce to $\cF^{(5)}$ and $\cF^{(6)}$.

We list the subleading Lagrangians we need below \cite{Chay:2002vy, Manohar:2002fd, Beneke:2002ph, Feldmann:2002cm, Pirjol:2002km, Beneke:2002ni, Bauer:2003mga}
\begin{eqnarray}
    \cL_{\xi \xi}^{(1)} &=& (\bar{\xi}_n W) i\Dslash^\perp_{us} \frac{1}{\bn \sdt \cP} \ (W^\dagger i\Dslash_c^\perp \hbns \ \xi_n) + (\bar{\xi}_n i\Dslash_c^\perp W) \frac{1}{\bn \sdt \cP} \ i\Dslash_{us}^\perp (W^\dagger \hbns \ \xi_n), \\
    \cL_{\xi \xi}^{(2)} &=& (\bar{\xi}_n W) i\Dslash^\perp_{us} \frac{1}{\bn \sdt \cP} \ i\Dslash_{us}^\perp \hbns \ (W^\dagger \xi_n) + (\bar{\xi}_n i\Dslash_c^\perp W) \frac{1}{(\bn \sdt \cP)^2} \ i \bn \sdt D_{us} \hbns \ (W^\dagger i\Dslash_c^\perp \xi_n), \\
    \cL^{(1)}_{\xi q} &=& \bar{\xi}_n \frac{1}{i\bn \sdt D_c} \ ig \Bslash_c^{\perp} W q_{us} + \hc,
\end{eqnarray}
where $ig \Bslash_c^\perp = [i\bn \sdt D_c, i\Dslash_c^\perp]$.

\subsection{NLO with $n \sdt (p+q)=0$}
\label{sec:NLO_n}

At NLO things become more interesting, because both collinear and ultrasoft fields are involved. To get the right power counting one of the outgoing gluons must be collinear and the other usoft. The only additional assumption that makes sense here is
\begin{equation}
    n \sdt (p+q) = 0,
\end{equation}
because $\bn \sdt p \sim 1 \gg \la^2 \sim \bn \sdt q$.

The axial current gives rise to
\begin{eqnarray}
    \lan \textrm{T} \big\{ \cJ^{(4)} \ i\cL^{(1)} \big\} + \cJ^{(5)}\ran &=& \tfrac{1}{2}i \bn \sdt (p+q) \ \lan \left( \textrm{T} \big\{n \sdt J_5^{(4)} \ i\cL^{(1)} \big\} + n \sdt J_5^{(5)} \right) \ran \\
    & \ & \hspace{-5.5cm} = \tfrac{1}{2}i \bn \sdt (p+q) \ \lan \bigg(\textrm{T} \Big\{ (
    -\bar{\xi}_n i \! \overleftarrow{\Dslash}_\perp^c \fr{1}{i\bn \sdt \overleftarrow{D}_c} \ \bnslash \ga_5 \fr{1}{i\bn \sdt D_c} \ i\Dslash_\perp^c \xi_n) \ i\cL^{(1)} \Big\} + \nonumber \\
    & \ & \hspace{-1.8cm} \Big( -2 \bar{\xi}_n i \! \overleftarrow{\Dslash}_\perp^c \fr{1}{i\bn \sdt \overleftarrow{D}_c} \ \ga_5 (W q_{us} + W \fr{1}{\bn \sdt \cP} \ i\Dslash_\perp^{us} \hbns \ W^\dagger \xi_n) + \hc \Big) \bigg) \ran. \nonumber
\end{eqnarray}

\begin{figure*}[h]
  \centerline{
   \mbox{\epsfxsize=3truecm \hbox{\epsfbox{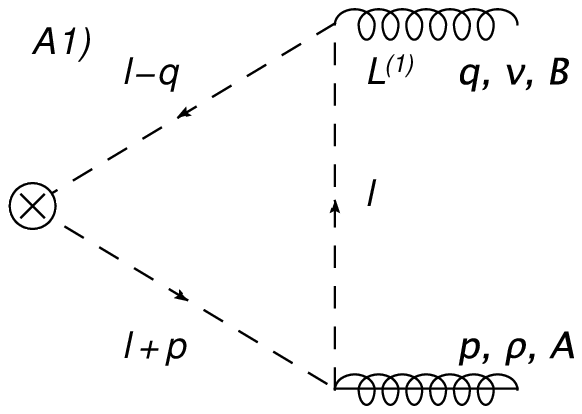}} } \quad
   \mbox{\epsfxsize=3truecm \hbox{\epsfbox{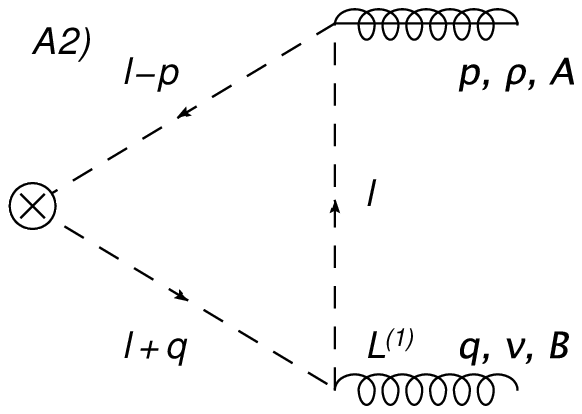}} } \quad
   \mbox{\epsfxsize=4truecm \hbox{\epsfbox{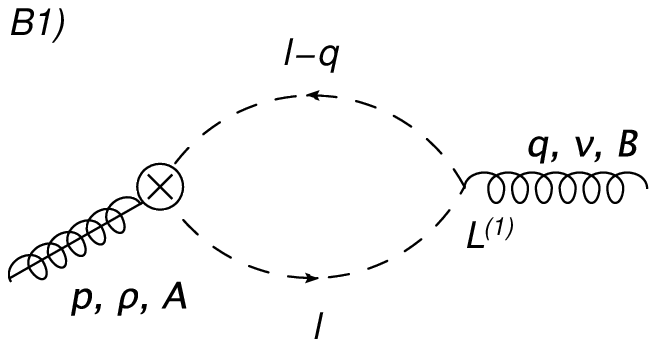}} } \quad
   \mbox{\epsfxsize=4truecm \hbox{\epsfbox{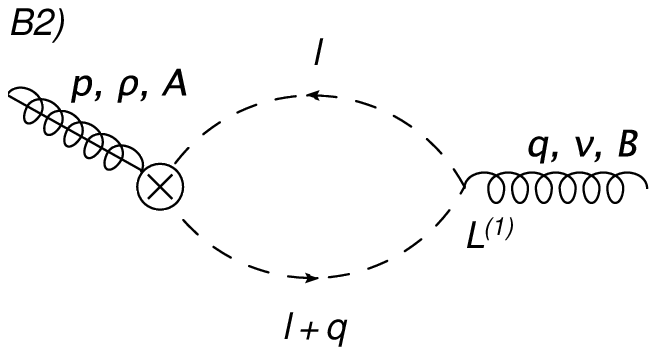}} }}
  \caption{Diagrams from $n \sdt J_5^{(4)}$ contributing to the NLO anomaly. The usoft gluon vertex is $\cL^{(1)}$}
   \label{fig:NLO1}
\end{figure*}

\begin{figure*}[h]
  \centerline{
   \mbox{\epsfxsize=4truecm \hbox{\epsfbox{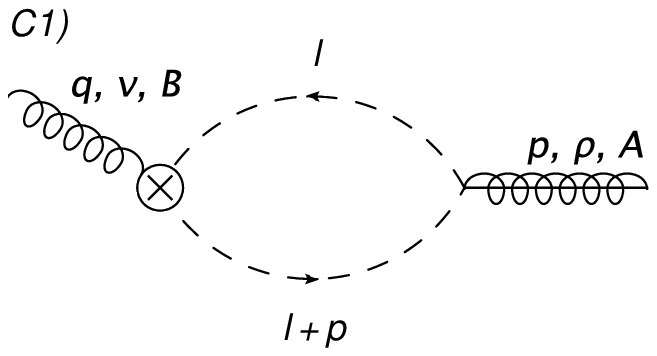}} } \quad
   \mbox{\epsfxsize=4truecm \hbox{\epsfbox{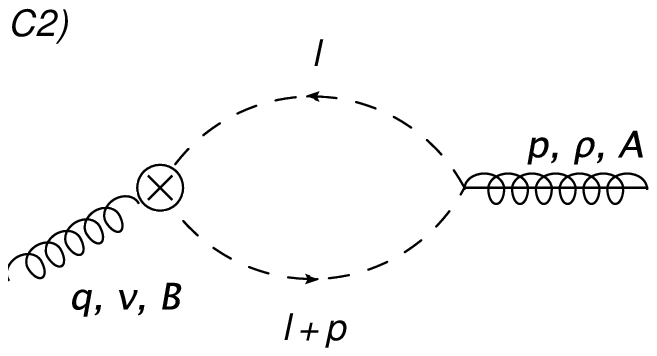}} } \quad
   \mbox{\epsfxsize=3truecm \hbox{\epsfbox{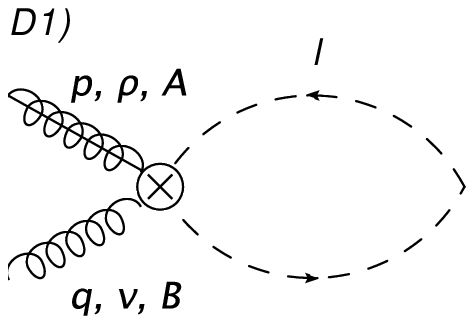}} } \quad
   \mbox{\epsfxsize=3truecm \hbox{\epsfbox{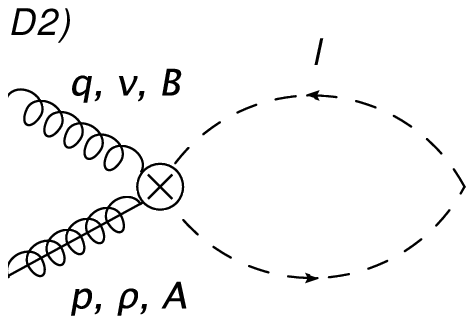}} }}
  \caption{Diagrams coming from $n \sdt J_5^{(5)}$ contributing to the NLO anomaly}
   \label{fig:NLO2}
\end{figure*}

Said in words, we get contributions from diagrams with $n \sdt J_5^{(4)}$, one $\cL^{(1)}$ vertex and $\cL^{(0)}$ vertices, and diagrams with $n \sdt J_5^{(5)}$ and only $\cL^{(0)}$ vertices. There is no term in $\cL^{(0)}$ for a $\perp$-polarized usoft gluon, so the usoft gluon has to come out of the $\cL^{(1)}$ vertex or the $n \sdt J_5^{(5)}$ current. This leads to the diagrams in figure \ref{fig:NLO1} and \ref{fig:NLO2}.
Obviously we do not have $\pqrn$ terms in this case, since one gluon is collinear and one usoft. As it turns out, we already computed all the necessary integrals in the LO $n \sdt (p+q)=0$ calculation. Here we have the additional simplification that we can generally drop $\bn \sdt q$ for the usoft momentum $q$, because $\bn \cdot q \ll \bn \cdot p, \bn \cdot l$.

The various diagrams lead to contributions
\begin{eqnarray}\label{NLO_contributions}
    A^{NLO}_{A1} + A^{NLO}_{A2} &=& 2 ig^2 \bn \sdt p \ \delta^{AB} \epsp_{\rho \nu} (1+\ve) (\tilde{I}_1 - \tilde{I}_2), \\
    A^{NLO}_{B1} + A^{NLO}_{B2} &=& 2 ig^2 \bn \sdt p \ \delta^{AB} \epsp_{\rho\nu} (1+\ve) \tilde{I}_{2A}, \\
    A^{NLO}_{C1} + A^{NLO}_{C2} &=& 2 ig^2 \bn \sdt p \ \delta^{AB} \epsp_{\rho\nu} (\tilde{I}_6 + \ve \tilde{I}_7), \\
    A^{NLO}_{D1} + A^{NLO}_{D2} &=& 0,
\end{eqnarray}
where $\tilde{I}_{2A}$ was defined in \eqref{integral_2_tilde}.
Adding the pieces we get
\begin{eqnarray}
    \lan \textrm{T} \big\{ \cJ^{(4)} \ i\cL^{(1)} \big\} + \cJ^{(5)}\ran &=& A^{NLO}_{A1} + A^{NLO}_{A2} + A^{NLO}_{B1} + A^{NLO}_{B2} + \nonumber \\
    & \ & A^{NLO}_{C1} + A^{NLO}_{C2} + A^{NLO}_{D1} + A^{NLO}_{D2} \nonumber \\
    &=& \fr{g^2}{8 \pi^2} \ \bn \sdt p \ n \sdt q \ \delta^{AB} \epsp_{\rho \nu}
    = \lan \cF^{(5)} \ran.
\end{eqnarray}
So the SCET anomaly equation is correct at NLO too.

\section{The SCET anomaly at NNLO}
\label{sec:SCET_anomaly_NNLO}

We will pursue our calculation to one higher order in $\lambda$. At this order the loop momentum can have either collinear or ultrasoft scaling.

Let us start by observing that there is a freedom in what you call label and residual momentum. Consequently SCET should be invariant under (for example) the following reparametrization
\begin{equation}
    \bn \sdt p_c \ra \bn \sdt p_c + \bn \sdt k, \qquad \bn \sdt p_r \ra \bn \sdt p_r - \bn \sdt k,
\end{equation}
where $p_c$ is the (collinear) label momentum, $p_r$ the (usoft) residual momentum and $k$ some constant usoft momentum. This ties together $\cF^{(4)}$ and $\cF^{(6)}$, basically predicting the latter. It is easy to check that \eqref{FFdual_at_lambda_4_2}+\eqref{FFdual_at_lambda_6} satisfies this. We will also calculate the loop diagrams, to verify the SCET anomaly equation at this order.
Since both gluons are collinear this time, either $\bn \sdt (p+q)=0$ or $n \sdt (p+q)=0$ can be imposed. We will restrict ourselves to the former. As can be seen from the expression for $\cF^{(6)}$ in \eqref{FFdual_at_lambda_6}, we need to keep the residual momentum components of the external momenta
\begin{equation}
    p = p_c + p_r = (0, \bn \sdt p_c, p_c^\perp = 0) + (n \sdt p_r, \bn \sdt p_r, p_r^\perp = 0)
\end{equation}
to get a non-vanishing expression. Our assumptions are
\begin{equation}
    \bn \sdt (p_c+q_c)=0, \qquad \bn \sdt (p_r + q_r)=0.
\end{equation}

The axial current gives
\begin{eqnarray}
    \lan \textrm{T} \big\{ \cJ^{(4)} \ \tfrac{1}{2} \ i\cL^{(1)} \ i\cL^{(1)} \big\} + \textrm{T} \big\{ \cJ^{(4)} \ i\cL^{(2)} \big\} + \textrm{T} \big\{ \cJ^{(5)} \ i\cL^{(1)} \big\} + \cJ^{(6)} \ran & \ & \nonumber \\
    & \ & \hspace{-11cm} = \tfrac{1}{2}i n \sdt (p+q) \ \lan \bigg( \textrm{T} \Big\{ \bn \sdt J_5^{(2)} \ \tfrac{1}{2} \ i\cL^{(1)} \ i\cL^{(1)} \Big\} + \textrm{T} \Big\{ \bn \sdt J_5^{(2)} \ i\cL^{(2)} \Big\} + \nonumber \\
    & \ & \hspace{-7.4cm} \textrm{T} \Big\{ \bn \sdt J_5^{(3)} \ i\cL^{(1)} \Big\} + \bn \sdt J_5^{(4)} \bigg) \ran \nonumber \\
    & \ & \hspace{-11cm} = \tfrac{1}{2}i n \sdt (p+q) \ \lan \bigg( \textrm{T} \Big\{ \bar{\xi}_n \bnslash \ga_5 \xi_n \ \tfrac{1}{2} \ i\cL^{(1)} \ i\cL^{(1)} \Big\} + \textrm{T} \Big\{ \bar{\xi}_n \bnslash \ga_5 \xi_n \ i\cL^{(2)} \Big\} + \nonumber \\
    & \ & \hspace{-7.3cm} 0 + (\bar{\xi}_n \bnslash \ga_5 W q_{us} + \hc ) \bigg) \ran.
\end{eqnarray}
We cannot make a diagram using $\bn \sdt J_5^{(4)}$ and only $\cL^{(0)}$ vertices, since the $\bn \sdt J_5^{(4)}$ current has a collinear and an usoft quark coming out of it. Therefore the only contributions come from $\bn \sdt J_5^{(2)}$.

There is a bit of a technical issue with loop integrals that we need to discuss here. At LO and NLO we had collinear loops for which we combined label and residual momenta $\sum_{l_c} \int \bard l_r \ra \int \bard l$. This was possible because these diagrams only involved $\bn \sdt l_c$, $l_c^\perp$ and $n \sdt l_r$. Now we can get diagrams that also depend on other components of the residual loop momenta e.g. $l_r^\perp$. This requires us to do $\int \bard l_r^\perp (...)$ separately, which yields zero for a collinear loop since all factors of $l_r^\perp$ are in the numerator. This is not the case when the loop momentum is usoft, because then $l_r^\perp$ appears in the denominator as well. Diagrams with collinear loops may still have non-zero contributions, for example $\sum_{l_c} \int \bard l_r \ \bn \sdt (l_r+p_r)(...) = \bn \sdt p_r \sum_{l_c} \int \bard l_r (...)$. Because we take the external $p_r^\perp = q_r^\perp = 0$, many diagrams vanish immediately. \\

\begin{figure*}[t]
  \centerline{
   \mbox{\epsfxsize=4truecm \hbox{\epsfbox{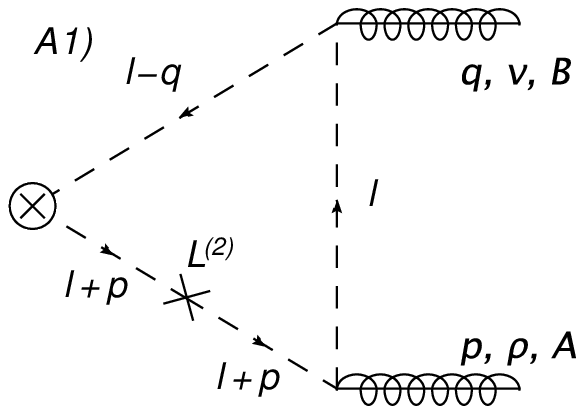}} } \qquad
   \mbox{\epsfxsize=4truecm \hbox{\epsfbox{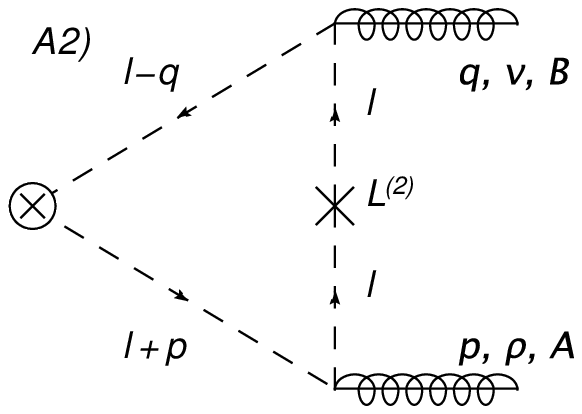}} } \qquad
   \mbox{\epsfxsize=4truecm \hbox{\epsfbox{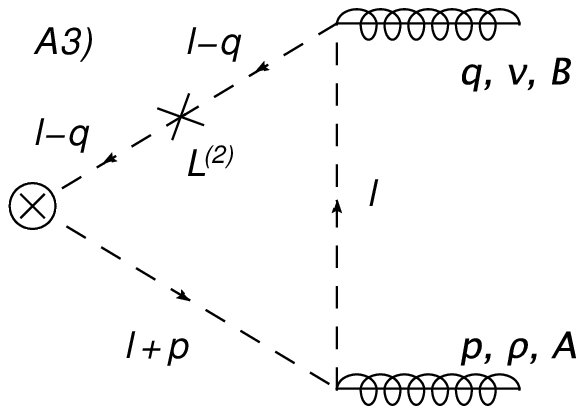}} }}
  \vspace{0.5cm}
  \centerline{
   \mbox{\epsfxsize=3.7truecm \hbox{\epsfbox{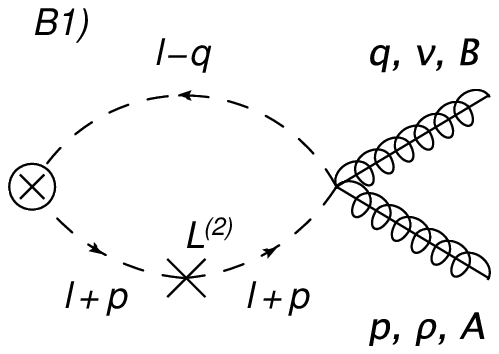}} } \qquad
   \mbox{\epsfxsize=3.7truecm \hbox{\epsfbox{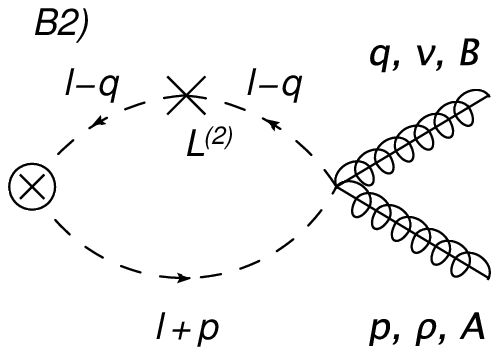}} } \qquad
   \mbox{\epsfxsize=4truecm \hbox{\epsfbox{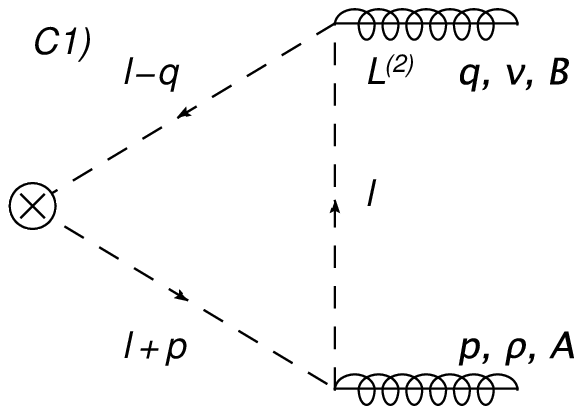}} }}
  \vspace{0.5cm}
  \centerline{
   \mbox{\epsfxsize=4truecm \hbox{\epsfbox{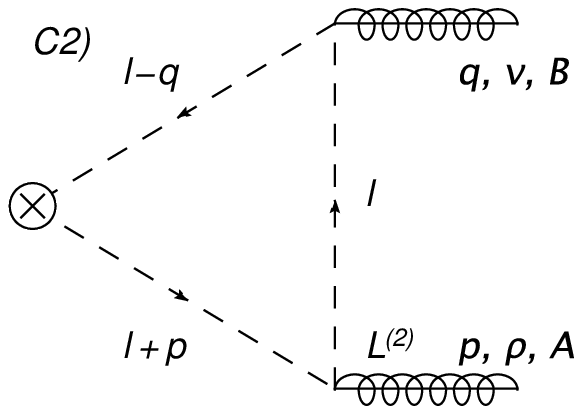}} } \qquad
   \mbox{\epsfxsize=3.7truecm \hbox{\epsfbox{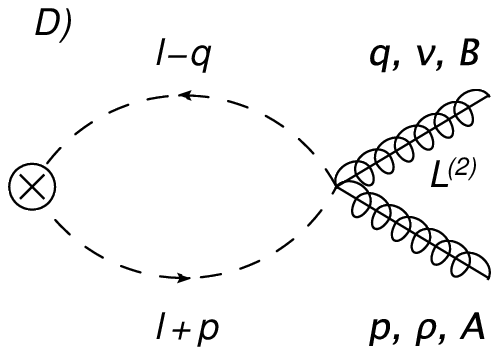}} } \qquad
   \mbox{\epsfxsize=4truecm \hbox{\epsfbox{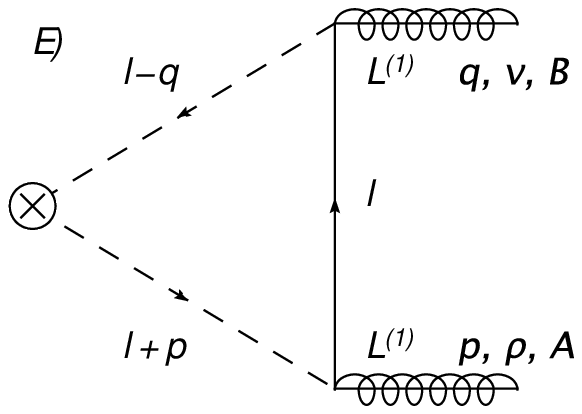}} }}
  \caption{Diagrams contributing to the anomaly at NNLO for $\bn \sdt (p+q)=0$}
   \label{fig:NNLO}
\end{figure*}
First of all we get the same diagrams as at LO, but with an extra $\cL^{(2)}$ insertion (fig. \ref{fig:NNLO}A1-B2)
\begin{multline}
    A^{NNLO}_{A1} + A^{NNLO}_{A2} + A^{NNLO}_{A3} + A^{NNLO}_{B1} + A^{NNLO}_{B2} = \\
    -2ig^2 n\sdt(p+q) \bn \sdt p_r \ \de^{AB} \epsp_{\rho \nu}
    \left(\hat{I}_1 + 2\ve \hat{I}_2 - (1+2\ve) \hat{I}_4 - \hat{I}_5 \right) + \pqrn,
\end{multline}
where
\begin{eqnarray}
    \hat{I}_1 = \int \bard l \fr{\bn \sdt (l+p) \ l_\perp^4}{\bn \sdt l \ (l-q)^2 \ l^2 \ (l+p)^4}, &\quad&
    \hat{I}_2 = \int \bard l \fr{l_\perp^4}{(l-q)^2 \ l^2 \ (l+p)^4}, \\
    \hat{I}_4 = \int \bard l \fr{\bn \sdt l \ l_\perp^4}{\bn \sdt (l+p) \ (l-q)^2 \ l^2 \ (l+p)^4}, \nonumber &\quad&
    \hat{I}_5 = \int \bard l \fr{\bn \sdt (l+p) \ l_\perp^2}{\bn \sdt l \ (l-q)^2 \ (l+p)^4}. \nonumber
\end{eqnarray}
We also get a triangle (fig. \ref{fig:NNLO}C1, C2) where one of the vertices is $\cL^{(2)}$ and a bubble diagram (fig. \ref{fig:NNLO}D) with a $\cL^{(2)}$ vertex
\begin{multline}
    A^{NNLO}_{C1} + A^{NNLO}_{C2} + A^{NNLO}_D = \\
    2ig^2 n \sdt (p+q) \bn \sdt p_r \ \delta^{AB} \epsp_{\rho\nu} (\ve \hat{I}_8 - (1+2\ve) \hat{I}_9 + \pqrn + \hat{I}_{10}),
\end{multline}
with
\begin{eqnarray}
    \hat{I}_8 = \int \bard l \fr{l_\perp^2}{(l-q)^2 \ l^2 \ (l+p)^2}, &\quad&
    \hat{I}_9 = \int \bard l \fr{\bn \sdt l \ l_\perp^2}{\bn \sdt (l+p) \ (l-q)^2 \ l^2 \ (l+p)^2}, \\
    \hat{I}_{10} = \int \bard l \fr{(\bn \sdt (l+p))^2}{\bn \sdt l \ (l-q)^2 \ (l+p)^2}.
\end{eqnarray}
Finally there is also a mixed collinear-usoft diagram (fig. \ref{fig:NNLO}E) with an usoft loop momentum
\begin{eqnarray}
    A^{NNLO}_E &=& -i g^2 n \sdt (p+q) \delta^{AB} \epsp_{\rho \nu} \int \bard l
    \fr{-\bn \sdt q}{-\bn \sdt q \ n \sdt (l-q)} \fr{n \sdt l}{l^2} \fr{\bn \sdt p}{\bn \sdt p \ n \sdt (l+p)} + \\
    & \ & \pqrn \nonumber \\
    &=& 0. \nonumber
\end{eqnarray}
In a naive calculation where you forget to treat $l$ as a residual momentum that is multipole expanded in the collinear propagators, you would misleadingly find a non-zero result for this diagram. \\

Rewriting the loop integrals using \eqref{simplify_integrals} and
\begin{equation}
    \fr{d}{d(n\sdt p)} \fr{1}{(l+p)^2} = -\fr{\bn (l+p)}{(l+p)^4}, \qquad
    \fr{d}{d(\bn\sdt p)} \fr{1}{(l+p)^2} = -\fr{n (l+p)}{(l+p)^4},
\end{equation}
we get
\begin{eqnarray}
    \hat{I}_1 &=& \hat{I}_2 + \hat{I}_5 - \hat{I}_8 - n\sdt p \ \tfrac{d}{d(n \cdot p)} \hat{I}_8, \\
    \hat{I}_4 &=& \hat{I}_2 - \hat{I}_8 + \hat{I}_9 - \bn \sdt p \ \tfrac{d}{d(\bn \cdot p)} \hat{I}_8.
\end{eqnarray}
This reduces the calculation to
\begin{eqnarray}
    \lan \textrm{T} \big\{ \cJ^{(4)} \ \tfrac{1}{2} \ i\cL^{(1)} \ i\cL^{(1)} \big\} + \textrm{T} \big\{ \cJ^{(4)} \ i\cL^{(2)} \big\} + \textrm{T} \big\{ \cJ^{(5)} \ i\cL^{(1)} \big\} + \cJ^{(6)} \ran & \ & \nonumber \\
    & \ & \hspace{-12cm} = A^{NNLO}_{A1} + A^{NNLO}_{A2} + \dots + A^{NNLO}_D + A^{NNLO}_E \nonumber \\
    & \ & \hspace{-12cm} = -2ig^2 n \sdt (p+q) \bn \sdt p_r \ \de^{AB} \epsp_{\rho\nu} \left(\ve \hat{I}_8 - n\sdt p \ \tfrac{d}{d(n \cdot p)} \hat{I}_8 + \nonumber
    (1+2\ve) \bn \sdt p \ \tfrac{d}{d(\bn \cdot p)} \hat{I}_8 - \tfrac{1}{2}\hat{I}_{10} \right) \nonumber \\
    & \ & \hspace{-12cm} \phantom{=} + \pqrn.
\end{eqnarray}
Working out $\hat{I}_{10}$, one finds zero. The remainder is fairly easy to evaluate if you take $\pqrn$ before performing the integral over Feynman parameters.
We find
\begin{eqnarray}
    \lan \textrm{T} \big\{ \cJ^{(4)} \ \tfrac{1}{2} \ i\cL^{(1)} \ i\cL^{(1)} \big\} + \textrm{T} \big\{ \cJ^{(4)} \ i\cL^{(2)} \big\} + \textrm{T} \big\{ \cJ^{(5)} \ i\cL^{(1)} \big\} + \cJ^{(6)} \ran & \ & \nonumber \\
    & \ & \hspace{-6 cm} = \fr{g^2}{8\pi^2} \ n \sdt (p+q) \bn \sdt p_r \ \delta^{AB} \epsp_{\rho\nu} = \lan \cF^{(6)} \ran.
\end{eqnarray}
Once again we see by direct computation that the SCET anomaly equation is correct.

\newpage

\section{Conclusions}
\label{sec:conclusions}

We derived the chiral anomaly equations in SCET up to NNLO by matching the QCD anomaly equation onto SCET, using the tree level relations between fields in QCD and SCET.
Gathering all the expressions, the anomaly equations up to NLO read\footnote{We only studied some terms of the NNLO anomaly equations and would need the higher order terms in \eqref{psi_in_SCET_full} to write down the full expression at NNLO.}
\begin{align}
    &\cJ^{(4)} = \cF^{(4)}, \\
    &\textrm{T} \big\{ \cJ^{(4)} \ i\cL^{(1)} \big\} + \cJ^{(5)} = \textrm{T} \big\{ \cF^{(4)} \ i\cL^{(1)} \big\} + \cF^{(5)},
\end{align}
where
\begin{eqnarray}
    \cJ^{(4)} &=& \tfrac{1}{2} n \sdt \partial \bigg[ \bar{\xi}_n \bnslash \ga_5 \xi_n \bigg] + \tfrac{1}{2} \bn \sdt \partial \bigg[ -\bar{\xi}_n i \! \overleftarrow{\Dslash}_\perp^c \fr{1}{i\bn \sdt \overleftarrow{D}_c} \ \bnslash \ga_5 \fr{1}{i\bn \sdt D_c} \ i\Dslash_\perp^c \xi_n \bigg] + \nonumber \\
    & \phantom{=} & \partial^\perp_\mu \bigg[ \bar{\xi}_n \ga_\perp^\mu \ga_5 \fr{1}{i\bn \sdt D_c} \ i\Dslash_\perp^c \hbns \xi_n + \hc \bigg], \\
    \cJ^{(5)} &=& \tfrac{1}{2} \bn \sdt \partial \bigg[ -2 \bar{\xi}_n i \! \overleftarrow{\Dslash}_\perp^c \fr{1}{i\bn \sdt \overleftarrow{D}_c} \ \ga_5 (W q_{us} + W \fr{1}{\bn \sdt \cP} i\Dslash_\perp^{us} \hbns \ W^\dagger \xi_n) \bigg] + \nonumber \\
    & \phantom{=} & \partial^\perp_\mu \bigg[ \bar{\xi}_n \ga_\perp^\mu \ga_5 (W q_{us} + W \fr{1}{\bn \sdt \cP} \ i\Dslash_\perp^{us} \hbns \ W^\dagger \xi_n) \bigg] + \hc, \\
    \cF^{(4)} &=& \fr{1}{16\pi^2} \ \epsp_{\mu\nu} \tr \big[ in \sdt D \ i \bn \sdt D_c \ iD_c^{\perp \mu} \ iD_c^{\perp \nu} \pm \textrm{permutations} \big], \\
    \cF^{(5)} &=& \fr{1}{8\pi^2} \ \epsp_{\mu\nu} \tr \big[ in \sdt D \ i \bn \sdt D_c \ iD_c^{\perp \mu} \ (W iD_{us}^{\perp \nu} W^\dagger) \pm \textrm{permutations} \big].
\end{eqnarray}
Here the sum over permutations means the sum of all different orderings of the operators multiplied by the sign of the permutation. Note that $\cF^{(4)}$ has an additional factor of $\tfrac{1}{2}$ compared to $\cF^{(5)}$, which comes from the redundant interchange\footnote{In our derivation of $\cF^{(4)}$ in \eqref{FFdual_at_lambda_4} we did not have this factor $\tfrac{1}{2}$, because we were distinguishing the outgoing gluons.} of the $iD_c^\perp$. The relevant terms of $\cL^{(1)}$ can be found at the end of sec. \ref{sec:match_QCD_anomaly_higher}.

By explicitly calculating one loop graphs we verified these equations.
Although we have not checked every possible term in the operator relations, our work suggests their correctness.
At LO the anomaly only involves collinear fields and SCET with only collinear fields is just a boosted version of QCD, so we expected agreement. It was not a priori clear how the anomaly equations would work out beyond that order.

Of course there are still higher orders to consider, which might be worth pursuing if one has in mind applying the anomaly in some specific process. $\cF$ contains terms up to order $\la^8$, whereas $\cJ$ has terms of arbitrary high order. This can be seen from replacing $1/(in \sdt D_c) \ra 1/(in \sdt D_c + W in \sdt D_{us} W^\dagger)$ in \eqref{psi_in_SCET_full} and expanding. If our result extends to higher orders, then certain matrix elements of the currents will have to cancel each other beyond $\la^8$.

\section{Acknowledgements}

I would like to thank Iain Stewart for many good discussions and helpful suggestions regarding this work.
This work was supported in part by funds provided by the U.S. Department of Energy (DOE) under cooperative research agreement DF-FC02-94ER40818.

\appendix

\section{Useful integrals}
\label{sec:useful_int}

All the loop integrals needed for our calculations are given below:
\begin{eqnarray}\label{loop_int}
    \int \bard^d l \fr{1}{(l^2-m^2)^\nu} &=&
    \fr{i(-1)^\nu}{16\pi^2} \ (1 + \log 4\pi \ \ve + \dots) \
    \fr{\Ga(\nu-2+\ve)}{\Ga(\nu)} \ (m^2)^{2-\nu-\ve} \\
    \int \bard^d l \fr{l_\perp^2}{(l^2-m^2)^\nu} &=&
    \fr{i(-1)^{\nu+1}}{16\pi^2} \ (1 + (\log 4\pi -1)\ve + \dots) \
    \fr{\Ga(\nu-3+\ve)}{\Ga(\nu)} \ (m^2)^{3-\nu-\ve} \\
    \hspace{-1.3 cm} \int \bard^d l \fr{1}{\bn \sdt (l+a) \ (l^2-m^2)^\nu} &=&
    \fr{i(-1)^\nu}{16\pi^2} \ (1 + \log 4\pi \ \ve + \dots) \
    \fr{\Ga(\nu-2+\ve)}{\Ga(\nu)} \ \fr{(m^2)^{2-\nu-\ve}}{\bn \sdt a} \\
    \hspace{-1.3 cm} \int \bard^d l \fr{l_\perp^2}{\bn \sdt (l+a) \ (l^2-m^2)^\nu} &=&  \fr{i(-1)^{\nu+1}}{16\pi^2} \ (1 + (\log 4\pi -1) \ve + \dots) \
    \fr{\Ga(\nu-3+\ve)}{\Ga(\nu)} \ \fr{(m^2)^{3-\nu-\ve}}{\bn \sdt a}
\end{eqnarray}
where $\int \bard^d l = \int d^d l (2\pi)^{-d}$ and $d = 4-2\ve$. Note that $l_\perp^2 = -\vec{l}_\perp^2$ when $g^{\mu \nu} = \textrm{diag}(+---)$.

\bibliography{anomalies}

\end{document}